\documentclass[aps,prl,showpacs,%
preprintnumbers,%
twocolumn,amsfonts,amssymb,superscriptaddress]{revtex4}
\usepackage{placeins}
\usepackage{graphicx}
\usepackage{amsmath}
\usepackage{times}
\usepackage{marvosym}
\begin{document}
\preprint{Physical Review Letters 95, 238701 (2005)} 
\title{Coevolutionary Dynamics: From Finite to Infinite Populations}
\author{Arne Traulsen}
\email{traulsen@fas.harvard.edu}
 \affiliation{Institut f{\"u}r Theoretische Physik und
Astrophysik, Christian-Albrechts Universit{\"a}t,
Olshausenstra{\ss}e 40, 24098 Kiel, Germany}
\affiliation{Program for Evolutionary Dynamics,
Harvard University,
One Brattle Square,
Cambridge, MA 02138, USA} 
 \author{Jens Christian Claussen}
 \affiliation{Institut f{\"u}r Theoretische Physik und
Astrophysik, Christian-Albrechts Universit{\"a}t,
Olshausenstra{\ss}e 40, 24098 Kiel, Germany}
\author{Christoph Hauert}
\affiliation{Program for Evolutionary Dynamics,
Harvard University,
One Brattle Square,
Cambridge, MA 02138, USA} 
\date{July 14 and September 15, 2005}
\begin{abstract}
Traditionally, frequency dependent evolutionary dynamics is described by deterministic replicator dynamics assuming implicitly infinite population sizes. 
Only recently have stochastic processes been introduced to study evolutionary dynamics in finite populations. However, the relationship between deterministic and stochastic approaches remained unclear. Here we solve this problem by explicitly considering large populations. In particular, we identify different microscopic stochastic processes that lead to the standard or the adjusted replicator dynamics. Moreover, differences on the individual level can lead to qualitatively different dynamics in asymmetric conflicts and, depending on the population size, can even invert the direction of the evolutionary process.
\end{abstract}
\pacs{
02.50.Le,		
05.45.-a, 		
87.23.-n, 		
89.65.-s 		
\hfill
DOI: 10.1103/PhysRevLett.95.238701
}

\maketitle

Darwinian evolution represents an intrinsically frequency dependent process. The fitness or reproductive output of an individual is not only linked to environmental conditions but also tightly coupled to the type and frequency of its competitors. Evolutionary game theory \cite{maynard-smith/price:1973,taylor/jonker:1978,maynard-smith:1982,hofbauer/sigmund:1998} has become a powerful framework to investigate the evolutionary fate of individual traits with differing competing abilities. Consider a population of two types $A$ and $B$. The fitness (or payoff) of the two types depends on their interaction partners and is determined by the payoff matrix
\begin{equation}
\begin{array}{c|cc} 
& A & B \\
\hline
A & a & b \\ 
B & c & d
\end{array}.
\end{equation}
Traditionally, the dynamics of such systems is investigated in the context of the well-known replicator equation \cite{taylor/jonker:1978,maynard-smith:1982,hofbauer/sigmund:1998}, a deterministic differential equation describing the change in frequency of the two (or more) types in infinite populations. 
For two types, this results in four basic scenarios of coevolutionary dynamics \cite{nowak/sigmund:2004}, while complex dynamics can arise in higher dimensions
\cite{sato/crutchfield:2002}.

In nature, however, populations are finite in size and the deterministic selection process is augmented and disturbed by stochastic effects and random drift. This has long been recognized by population geneticists and goes back to the seminal work by Wright \cite{wright:1931} and Fisher \cite{fisher:1930}. Assuming a finite but constant population size the balance between selection and drift can be described by the Moran process \cite{moran:1962}. The microscopic dynamics consists of three simple steps: (i) \emph{selection}, an individual is randomly selected for reproduction with a probability proportional to its fitness; (ii) \emph{reproduction}, the selected individual produces one (identical) offspring; (iii) \emph{replacement}, the offspring replaces a randomly selected individual in the population.

The Moran process allows to derive the fixation probability of mutant genes or investigate the effect of population structures on the fixation probability \cite{lieberman/nowak:2005}. Originally, the Moran process was formulated in a frequency independent setting where the fitness of an individual is genetically determined and remains unaffected by interactions with other individuals 
as in Ref.\ \cite{kessler/levine:1998}. Only recently, the frequency dependent approach of evolutionary game theory and the Moran process have been successfully combined in order to investigate the evolutionary dynamics in finite populations \cite{nowak/fudenberg:2004, taylor/nowak:2004}. The fitness of an individual now comprises two components: the frequency independent baseline fitness which is associated with genetic predisposition and the frequency dependent contribution from interactions with other members of the population.

Thus, evolutionary dynamics can be described by a continuous deterministic replicator equation or by a stochastic microscopic description of a birth-death process such as the Moran process. So far, the relation and transition between the two approaches remained unclear. 
We show that different replicator dynamics are associated with different microscopic processes and moreover that the dynamics derived for infinite populations may undergo qualitative changes in finite populations.

If every individual interacts with a 
representative sample of the population, the average 
payoff of $A$ and $B$ individuals will be determined
by the fraction of coplayers of both types. Excluding self interactions,
this leads to the payoffs  
\begin{eqnarray}
\label{payoffeq}
\pi^A_i & = & \frac{a \, (i-1)+b(N-i)}{N-1} \\ \nonumber
\pi^B_i & = & \frac{c \, i+d(N-i-1)}{N-1},
\end{eqnarray}
where $i$ is the number of $A$ individuals and $N$
is the population size. 

The effective reproductive fitness of an individual of type $k$ is then given by $1-w+w \pi^k_i$ where $w$ determines the relative contributions of the baseline fitness, which is conveniently normalized to one, and the payoffs result from interactions with other individuals \cite{nowak/fudenberg:2004}. Thus, $w$ corresponds to the strength of selection acting on the game under consideration. Note that any $w<1$ can be mapped to a system with a different payoff matrix and $w=1$ \cite{claussen/traulsen:2005}.
For the Moran process, 
the probability that the number of $A$ individuals increases from $i$ to $i+1$ is
\begin{equation}
\label{transmatMoranA}
 T^+(i)  =  \frac{ {1-w}+ w\,\pi^A_i      }{{1-w}+ w\, \langle \pi_i \rangle}   \frac{i}{N} \frac{N-i}{N},  
\end{equation}
whereas it decreases from $i$ to $i-1$ with probability
\begin{equation}
\label{transmatMoranB}
 T^-(i)  =  \frac{ {1-w}+ w\, \pi^B_i      }{{1-w}+w\, \langle \pi_i \rangle}  
\frac{i}{N} \frac{N-i}{N}.
\end{equation}
$\langle \pi_i \rangle = \left( \pi^A_i \,i + \pi^B_i\,(N-i)\right)/N$ is the average
payoff in the population. 

Note that the selection mechanism in the Moran process requires perfect global information on the current state of the population. This is a very strong requirement and in many situations undesirable. Therefore, we propose an alternative formulation for the microscopic processes entirely based on local information: In each time step, a randomly chosen individual $b$ compares its payoff to the payoff of
another randomly chosen individual $a$. It switches to the other's strategy
with probability 
\begin{equation}
p = \frac{1}{2} + \frac{w}{2} \frac{\pi_a-\pi_b}{\Delta \pi_{\rm max}},
\end{equation}
where $\Delta \pi_{\rm max}$ is the maximum possible payoff 
difference and $0 < w \leq 1$ measures the strength of selection. 
Note that in contrast to the Moran process, the local update is invariant under linear rescaling of the payoff matrix.
The transition matrix for the number of A individuals in this process $i$
is then given by
\begin{eqnarray}
\label{transmatLHUR}
T^+(i)  & =  & \left( \frac{1}{2} + \frac{w}{2} \frac{\pi^A_i -\pi^B_i}{\Delta \pi_{\rm max}} \right)  \frac{i}{N} \frac{N-i}{N}   \nonumber \\
T^-(i)   &  = & \left( \frac{1}{2} + \frac{w}{2} \frac{\pi^B_i-\pi^A_i}{\Delta \pi_{\rm max}} \right) \frac{i}{N} \frac{N-i}{N} .
\end{eqnarray}
In both process, the number of A individuals remains constant with probability
 $ T^0(i)=1 - T^+(i) -T^-(i)$. Further, the states $i=0$ and $i=N$ are absorbing.

We can directly compute the probability that a single A
individual fixates in the population,
$\phi_A$. In general, this probability is given by \cite{karlin/taylor:1975}
\begin{equation}
\phi_A = \frac{1}{
1+ \sum_{k=1}^{N-1} \prod_{i=1}^k \frac{T^-(i)}{T^+(i)}}.
\label{fixprobMoran}
\end{equation}
In the limit of weak selection, $ w \ll 1$, the fixation probabilities can be computed analytically \cite{weakselection}. 
The fixation probability for the Moran process is higher 
than for the local update mechanism if $\Delta \pi_{\rm max} >2$.

The stochastic process can be formulated in terms of the master equation
\cite{kampen:1992,schuster:2002}
\begin{eqnarray}
 P^{\tau+1}(i) - P^{\tau}(i)
&=&
\hphantom{+}
P^{\tau}(i-1)  T^+(i-1)
-P^{\tau}(i)  T^- (i)
 \nonumber \\ \nonumber
&  &+  P^{\tau} (i+1)  T^- (i+1)
-P^{\tau}(i)  T^+ (i)
\end{eqnarray}
where $P^{\tau}(i)$ is the probability that
the system is in state $i$ at time $\tau$. 
Introducing the notation
$x=i/N$, $t=\tau/N$ and the 
probability density $\rho(x,t) = N\, P^{\tau}(i)$ 
yields
\begin{eqnarray}
 \rho\left(x,t+ N^{-1} \right) 
 -  \rho\left(x,t\right) \!
&=&
 \rho\left(x-N^{-1},t\right)   T^+(x-N^{-1})
\nonumber \\ \nonumber &&
\!\!\!\!\!
 + \rho\left(x+N^{-1},t\right)  T^- (x+N^{-1})
 \nonumber \\ \nonumber &&
\!\!\!\!\!
 -\rho\left(x,t\right)  T^- (x)
 -\rho\left(x,t\right)  T^+ (x).
\end{eqnarray}
For $N \gg 1$, the probability densities and
the transition probabilities are expanded in a Taylor series at $x$ and $t$. 
Neglecting higher order terms in $N^{-1}$ we get \cite{kampen:1992}
\begin{equation}
\label{FPE}
\frac{d}{d t} \rho(x,t) = - \frac{d}{dx} \left[ a(x) \rho(x,t) \right] + \frac{1}{2} \frac{d^2}{dx^2} \left[ b^2(x)  \rho(x,t) \right],
\end{equation}
with $a(x)\!=\! T^+\!(x)-T^-\!(x)$
and $b(x) \!= \!\sqrt{\frac{1}{N}\! \left[T^+\!(x)+T^-\!(x) \right]}$.
Note that, for large but finite $N$, this equation has the form of a Fokker-Planck 
equation. 
Since the internal noise is not correlated in time as subsequent
update steps are independent, the It{\^o} calculus
\cite{kampen:1992} can be applied to derive 
a Langevin equation
\begin{equation}
\label{langevin}
\dot x = a(x) + b(x) \xi,
\end{equation}
where $\xi$ is uncorrelated Gaussian noise
and $b(x)=0$ for $x=0$ and $x=1$. 
The multiplicative noise term leaves the absorbing nature of the boundaries unaffected. 
Qualitatively similar results are obtained by introducing noise in the payoff matrix
\cite{fudenberg/harris:1992,imhof:2005}. In contrast, to account for stochastic effects
through additive noise \cite{foster/young:1990, traulsen/schuster:2004} is problematic
because of the boundaries of $x$. 

For $N \to \infty$, the diffusion term $b(x)$ vanishes with $\frac{1}{\sqrt{N}}$ and
a deterministic equation is obtained.
For the Moran process, this yields  
\begin{eqnarray}
\label{ARD}
\dot x & = &  \lim_{N \to \infty} \left( \frac{ \pi^A_i -\pi^B_i }{\Gamma+\langle \pi_i \rangle} \frac{i}{N}\frac{N-i}{N} \right)  \\ \nonumber
& = &  x \left(\pi^A(x) -\langle \pi(x) \rangle\right)\frac{1}{\Gamma+\langle \pi(x) \rangle}, 
\end{eqnarray}
where $\pi^A(x)=x \, a+(1-x) b$, $\pi^B(x)=x\, c + (1-x) d$, and 
$\langle \pi(x) \rangle = x \,\pi^A(x) +(1-x)\pi^B(x)$.
$\Gamma=\frac{1-w}{w}$ is essentially the baseline fitness. 
Equation (\ref{ARD}) is the adjusted replicator dynamics introduced in 
Ref.\ \cite{maynard-smith:1982}; an alternative derivation
for imitation dynamics is given in \cite{hofbauer/schlag:2000}.

For the local update mechanism, we find
\begin{eqnarray}
\dot x & = &  
 \lim_{N \to \infty} \left(  w  \frac{\pi^A_i-\pi^B_i}{\Delta \pi_{\rm max}} \frac{i}{N}\frac{N-i}{N} \right) \nonumber \\
& = & \kappa x  \left(\pi^A(x) - \langle \pi(x) \rangle \right),
\label{RD}
\end{eqnarray}
where $\kappa= \frac{w}{\Delta \pi_{\rm max}}$ is a constant factor influencing the timescale only. 
Equation (\ref{RD}) is the standard replicator equation and represents the traditional approach to evolutionary dynamics in infinite populations \cite{hofbauer/sigmund:1998}. The difference between the two dynamics amounts to a dynamic rescaling of time which leaves the fixed points unchanged. Nevertheless, the differences in the microscopic updating can give rise to substantial differences in the macroscopic dynamics. 
To illustrate this, let us consider two famous examples: the Prisoner's Dilemma \cite{axelrod/hamilton:1981} and Dawkins Battle of the Sexes \cite{dawkins:1976}.

\begin{figure}[htbp]
\includegraphics[angle=270,totalheight=7.2cm]{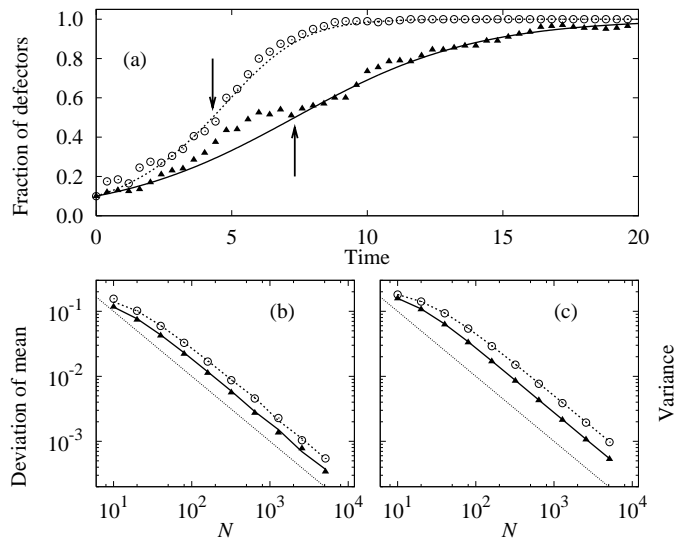}
\caption{ 
Prisoners Dilemma: (a) Approach to the Nash equilibrium from an initial state with 10\% defectors for simulations of the Moran process 
($\circ$) and the local update rule ($\blacktriangle$) with $N=200$ as compared to numerical solutions of the corresponding adjusted replicator dynamics (dashed line) and the standard replicator dynamics 
(solid line), respectively. The dynamics based on the Moran process converges much faster to the equilibrium state $x^\ast=1$. 
(b), (c) Convergence of the stochastic simulations ($\circ, \blacktriangle$) and the corresponding Langevin equations 
(dashed and solid lines) to the deterministic adjusted and standard replicator dynamics for increasing population sizes $N$. 
(b) Deviations of the mean fraction of defectors 
at time $t_1$ (depicted in (a) by the arrows) where the corresponding (adjusted) replicator dynamics predicts $x(t_1)=0.5$;
(c) same for the variance of the mean fraction of defectors. The results for the simulations and the Langevin equation are in excellent agreement. 
Both the deviations and the variance converge to the deterministic dynamics with $1/N$ (dotted line),
as expected ($b=1$, $c = 0.5$, $w = 0.9$, time scale for simulations $1/N$,
averages over $10^6$ realizations).}
\label{pdcomparison}
\end{figure}

{\em The Prisoner's Dilemma}
describes the problem of cooperation where two players simultaneously decide whether to cooperate (C) or defect (D). Cooperation incurs costs $c$ but produces a benefit $b$ to the other player ($b>c$) whereas defection neither costs nor benefits anyone. In this game defection is dominant because defectors are better off no matter what the other player does. Thus, rational players end up with nothing rather than earning the more favorable payoff for mutual cooperation 
--- hence the dilemma. The Prisoner's Dilemma is determined by the payoff matrix
\begin{equation}
\begin{array}{c|cc} 
& D & C \\
\hline
D & c & b+c \\ 
C & 0 & b \\ 
\end{array}
\label{pdpayoff}
\end{equation}
where we added $c$ in order to avoid negative payoffs. 
Figure \ref{pdcomparison} compares the evolution from a state with a small fraction of defectors into the evolutionary end state with defectors only. 
The standard replicator dynamics for
the Prisoner's Dilemma is given by
$\dot x = w \frac{c}{b+c} \, x (1-x)$, where $x$ is
the fraction of defectors.
The solution yields 
\begin{equation}
x(t) = x_0 \left[ x_0+(1-x_0) e^{-wtc/(b+c)} \right]^{-1}, 
\end{equation}
with $x(0)=x_0$ and $x(t\to \infty) = x^\ast =1$.
The adjusted replicator dynamics can be solved only
numerically for the given payoff matrix and general $w$. 
For $w > 0$, Fig.\ \ref{pdcomparison} illustrates that
the convergence to $x^\ast=1$ is faster for the adjusted replicator dynamics
as compared to the standard replicator dynamics.

{\em Dawkins Battle of the Sexes} \cite{dawkins:1976,bots} is a cyclic game
referring to asymmetric conflicts in parental care: If males (\Male) are philanderers ($B_{\text{\Male}}$), it pays for females (\Female) to be coy ($A_{\text{\Female}}$), insisting on a long courtship period to make males invest more in the offspring. However, once most males are faithful ($A_{\text{\Male}}$), fast females are favored ($B_{\text{\Female}}$) avoiding the costs of courtship. Subsequently, the male investment into the offspring is no longer justified, philanderers are again favored ($B_{\text{\Male}}$), and the cycle continues. This is characterized by the payoff 
matrix
\begin{equation}
\begin{array}{c|cc} 
& A_{\text{\Female}} & B_{\text{\Female}} \\
\hline
A_{\text{\Male}} & (+1,-1) & (-1,+1) \\ 
B_{\text{\Male}} & (-1,+1) & (+1,-1) \\ 
\end{array}
\end{equation}
where the first element is the payoff of the males and the
second element is the payoff of the females \cite{maynard-smith:1982}. 
Note that this game is also called 'Matching Pennies' \cite{weibull:1995,traulsen/schuster:2004}. 

\begin{figure}[htbp]
\includegraphics[totalheight=7.5cm]{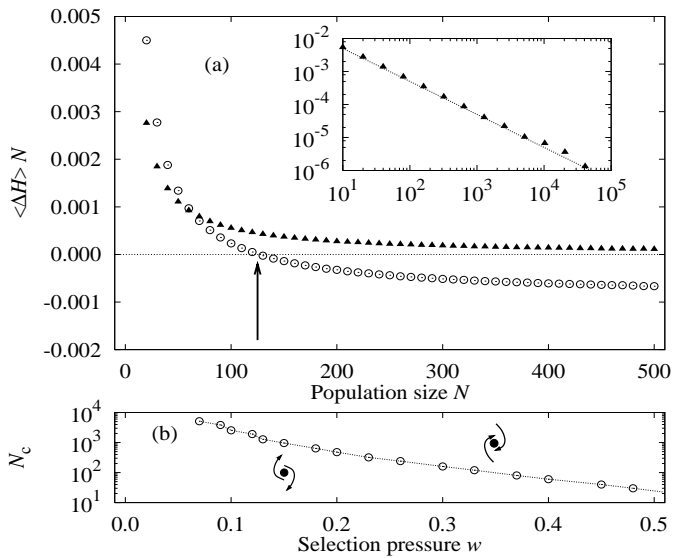}
\caption{Battle of the sexes: (a) Average drift in a finite population for different population size $N$ and fixed $w=0.3$
for the local update mechanism ($\blacktriangle$) and the Moran process ($\circ$).
For $\langle \Delta H \rangle <0$, the system spirals to the Nash equilibrium $\bf q$;
for $\langle \Delta H \rangle >0$, it spirals towards the absorbing boundaries. 
For the local update process, $\Delta H \to 0$ with $1/N$ (see inset),
which hints at the constant of motion in the limit $N \to \infty$ (see text).
For the Moran process, the system drifts towards the boundaries below 
a critical population size $N_c$ (marked by the arrow) but approaches $\bf q$ for $N>N_c$. (b) The critical population size $N_c$ (where 
$\langle \Delta H \rangle \approx 0$) decreases as a function of the selection pressure $w$  ($N=N_{\text{\Female}}=N_{\text{\Male}}$, averages over $10^7$ realizations). 
}
\label{fignum}
\normalsize
\end{figure}

The dynamics of Dawkins Battle of the Sexes is qualitatively different for the adjusted and the standard replicator dynamics \cite{maynard-smith:1982}. Comparisons of the deterministic dynamics to the corresponding stochastic process in finite populations reveal further interesting differences.
For the standard replicator dynamics $H=-x(1-x)y(1-y)$, where $x$ is the fraction of faithful males and $y$ the fraction of coy females, is a constant of motion which measures the distance from the Nash equilibrium ${\bf q}=(\frac{1}{2},\frac{1}{2})$. 
Consequently, $\bf q$ is a neutrally stable fixed point surrounded by periodic orbits. 
Figure \ref{fignum} shows that this is a spurious result valid only in the limit $N=N_{\text{\Female}}=N_{\text{\Male}}\to\infty$. For any finite $N$, $\dot H>0$ holds and every trajectory spirals away from $\bf q$. Note that in this case $\dot H$ is approximated by $\langle\Delta H\rangle=\langle H(t+1)-H(t)\rangle$, 
which is calculated for a single stochastic update step and averaged over many realizations. In contrast, the adjusted replicator dynamics leads to $\dot H\leq 0$, where equality holds only at the Nash equilibrium $\bf q$. $H$ represents a Lyapunov function with $\bf q$ as the unique and asymptotically stable interior fixed point. Interestingly, for finite $N$, i.e. for the Moran process, the drift towards $\bf q$ changes sign: 
Below a critical population size $N_c(w)$, $\bf q$ is unstable and the trajectories spiral towards the boundary.
However, for $N>N_c(w)$, $\bf q$ becomes stable, cf.\ Fig.\ \ref{fignum}. 
Similar qualitative changes based on population sizes have recently been reported in 
Refs.\ \cite{taylor/nowak:2004,imhof/nowak:2005}.

{\em In conclusion,} we presented a mathematically consistent transition from descriptions of the microscopic processes relevant for individual based simulations over stochastic approximations of the dynamics in finite populations to a deterministic mean field theory governed by replicator dynamics. In particular, we have shown that the intrinsic stochasticity arising in finite populations
can be captured by a Langevin term in the replicator dynamics. This
 leads naturally to absorbing boundaries in the resulting stochastic differential equation. 
 Both, the microscopic processes and their stochastic approximation converge with $1/N$
 to the solution of the corresponding replicator equation.

The frequency dependent Moran process \cite{nowak/fudenberg:2004} is intimately connected to the adjusted replicator dynamics \cite{maynard-smith:1982}. Conversely, the proposed local update rule corresponds to a finite population equivalent of the standard replicator dynamics \cite{hofbauer/sigmund:1998}. While the qualitative dynamics of the two approaches is the same, quantitative differences vanish only upon non-linear rescaling of time. However, for interacting (sub)populations, such rescaling is no longer possible and can give rise to entirely different qualitative dynamics. 
In particular, for Dawkins Battle of the Sexes, we have shown that, for the Moran process, the stability of the mixed Nash equilibrium depends on the population size such that it becomes unstable below a critical size and no mixed population can persist. 

John Maynard Smith asked whether the adjusted or the standard replicator dynamics is more appropriate to describe evolutionary changes \cite{maynard-smith:1982}. Here we have shown that this is fully determined by the underlying microscopic processes.

\bibliographystyle{plain}

\end{document}